\documentclass[a4paper,11pt, table]{article}

\usepackage{latexsym}
\usepackage{amssymb}
\usepackage{graphicx}
\usepackage{amsmath}
\usepackage{amsthm}
\usepackage{mathrsfs}
\usepackage{natbib}
\usepackage{url}
\usepackage{multirow}
\usepackage{array}
\usepackage{amsfonts}
\usepackage[hang,scriptsize]{caption}
\usepackage{hyperref}
\usepackage{fancyhdr}
\usepackage{floatrow, tabularx, makecell, booktabs}
\usepackage{enumerate}
\usepackage{subcaption}
\usepackage[a4paper,top=3cm,bottom=3cm,left=3cm,right=3cm]{geometry}
\usepackage[ruled, vlined, linesnumbered]{algorithm2e}  

\newtheorem{theorem}{Theorem}[section]
\newtheorem{definition}[theorem]{Definition}

\newtheorem{remark}[theorem]{Remark}

\newcommand\ind[1]{\mathbf{1}_{#1}}

\SetKwInOut{Parameter}{Global parameters}

\def\Xset{\ensuremath{\mathcal X}} 
\def\R{\ensuremath{\mathcal R}} 
\def\S{\ensuremath{\mathcal S}} 
\def\x{\ensuremath{\mathbf x}} 
\def\r{\ensuremath{\mathbf r}} 
\def\rowHRul{6cm}

\setcellgapes{3pt}

\title{ESG investments: Filtering versus machine learning approaches}
\author{Carmine de Franco, PhD\footnote{Portfolio Manager at Ossiam, carmine.de-franco@ossiam.com} \and 
		Christophe Geissler\footnote{CEO at Advestis, cgeissler@advestis.com} \and 	
		Vincent Margot\footnote{PhD Candidate, vmargot@advestis.com} \and 
		Bruno Monnier, CFA\footnote{Portfolio Manager at Ossiam, bruno.monnier@ossiam.com}}
\date{22 Oct 2018}

\begin{document}
\frenchspacing

\maketitle

\begin{abstract}
We designed a machine learning algorithm that identifies patterns between ESG profiles and financial performances for companies in a large investment universe. The algorithm consists of regularly updated sets of rules that map regions into the high-dimensional space of ESG features to excess return predictions. The final aggregated predictions are transformed into scores which allow us to design simple strategies that screen the investment universe for stocks with positive scores. By linking the ESG features with financial performances in a non-linear way, our strategy based upon our machine learning algorithm turns out to be an efficient stock picking tool, which outperforms classic strategies that screen stocks according to their ESG ratings, as the popular best-in-class approach. Our paper brings new ideas in the growing field of financial literature that investigates the links between ESG behavior and the economy. We show indeed that there is clearly some form of alpha in the ESG profile of a company, but that this alpha can be accessed only with powerful, non-linear techniques such as machine learning.
\end{abstract}
\medskip

\medskip

\section{Introduction}\label{intro}
\numberwithin{equation}{section}

The relationship between corporate social (CSP) and financial performances (CFP) is fairly an old theme in the economic research. In its earlier stages, it has met quite deep skepticism and critics: Nobel prize-winning economist Milton Friedman wrote in the New York Times Magazine, back in the 1970', that \emph{"... there is one and only one social responsibility of business - to use its resources and engage in activities designed to increase its profits so long as it stays within the rules of the game, which is to say, engages in open and free competition without deception or fraud...."} \citep{Friedman1970}. 

But the number of studies that highlights positive, or at least non-negative, relationship between social and financial performances has grown significantly since then, probably beginning with the initial work by \cite{BragdonMarlin1972} on the link between environmental virtue and financial performance. Fifty years later the context has completely changed. The number of proponents of social and, more broadly, ESG integration in both corporate management and investors' choices has grown exponentially. And so has the number of financial products, funds and ETFs, that offer ESG versions of a large panel of investment strategies (mainly on equity and bonds). 

The current approach now seems completely opposed to Friedman's one, and the most recent empirical literature highlights the link between ESG performance and alpha \citep{ChongPhillips2016,Giese2016, Zoltan2016}.
Nonetheless, the question regarding the relationship between CSP and CFP remains unanswered. Reviews of published paper (meta-analysis) highlight that the majority of empirical studies published on this theme reports non-negative or small positive relationship between CSP and CFP (see for example \cite{Orlitzky2003}, \cite{AlloucheLaroche2005}, \cite{Wu2006}, \cite{VanBeurdenGossling2008}, \cite{Margolis2009}, \cite{Friede2015}). Other researchers take a more optimistic view and report significant relationship between CSP and CFP \citep{PeirisEvans2010,Filbeck2014, IndraniClayman2015} or at least that CSP is not detrimental to CFP as long as one manages to build the portfolio with care, even if there is no clear value added in ESG integration \citep{KurtzDiBartolomeo2011}.

Although we do not share the very optimistic, and mostly overstated, enthusiasm about the direct relationship between ESG and financial performance, we do believe that there is a strong relationship between ESG and sustainability of corporate business. Therefore, ESG has an impact on financial performances and risks, but this does not come linearly.

\noindent We welcome the efforts that investors are undertaking to include ESG criteria into their portfolio choices, and we clearly hope that this will trigger economic and cultural changes in corporate management. At the same time, we remain skeptic in front of the far too flaunted capability of basic ESG ratings to act as an alpha generator in a portfolio.

It remains true however that the very large set of ESG data, reports and analysis can contain useful information related to the strengths and weaknesses of corporations. Unfortunately, ESG ratings are, by construction, a composite measure that dramatically reduces this rich set of information. 

Our contribution to the growing literature on this topic is to show that, empirically, there is no value added in portfolios based on simple ESG screenings. Although it usually comes with no harm to the performance, we do not find any alpha in such approaches. However, by recognizing the intrinsic value of the large panel of ESG indicators that are aggregated to form the ESG ratings, we show that it is possible to extract value from them, which, in turns, translates into real alpha. By exploring large data sets of specific ESG indicators, we are able to identify those who really have an impact on corporate financial performances. In a simplified example, we can agree on the fact that for a company in the utility sector, most likely, the environmental performance can be a discriminating criterion for financial performance; at the same time, governance can play an important role if we compare a utility company in Europe with one in an emerging country. Similarly, direct carbon emissions for banks are probably not as relevant as the exposures of these banks, through loans, to highly polluting companies would be. In short, aggregate measures as ESG ratings lose valuable information contained in the ESG indicators, which therefore lower their predictive power.

Searching for interesting patterns between specific ESG indicators and financial performance for a large set of companies remains out of reach for the standard tools available to econometricians. This search takes place in a very high-dimensional space and is not oriented by some a priori on these ESG features. To deal with this complexity, we developed a machine learning algorithm that allows us to identify features and patterns that are relevant to explain the link between CSP and CFP. The algorithm maps the regions in our high-dimensional space of ESG features that have been consistently associated with outperformance or underperformance. In the econometric parlance, we look at those regions for which the conditional expectation of each stock's forward return is statistically positive (or negative), given that its relevant ESG features fall in these regions. We say that these relevant ESG features "activate" the region. By observing the ESG features we then obtain a significant signal regarding the future financial performance of the stock.

This identification is done with a set of rules that take the form of \emph{If-Then} statements. The \emph{If} statement identifies the region in the ESG space: in other words, the values that some ESG features have to take in order to activate the rule. The \emph{Then} statement produces a prediction of the excess return, over the benchmark, that we can expect from a stock whose ESG features fall in that region. The final prediction is the aggregation of the predictions made by these rules and is transformed into a score $\{-1, 0, +1\}$. We therefore focus on the sign of the prediction of excess return rather than on its value. This usually makes the estimation more robust. The aggregation method mimics a panel of experts, each of which is expert on a very specific ESG feature (ex. Environment, Independence of the Board, ESG Reporting Verification, Employee Incidents, etc.) and makes a prediction given the ESG behavior of the company. When the aggregated prediction is close to zero, i.e the panel of experts is split between optimistic and pessimist forecasters, the final prediction is set at $0$. The algorithm is regularly trained over time so that it can react and readjust to the new observed data.  

The algorithm is used to design a very simple strategy that screens the investment universe and selects all stocks with a positive score. The resulting portfolio is compared with a classic ESG best-in-class portfolio, which consists of all stocks in the investment universe whose ESG ratings are above a given threshold within their peer groups. Our empirical results show that the simple machine learning screened portfolio significantly outperforms the ESG best-in-class approach and the benchmark. 

This is in line with the economic belief that ESG data is valuable information to assess financial performance, but also confirms that aggregated ESG ratings are not suited to distinguish between outperformers and underperformers over the long run. Even if perfect distinction is out of reach, our results clearly point out the fact that there is alpha in the granular ESG data, but the relation between ESG and financial performance is definitely not linear. Furthermore, the predictive power of the scores vanishes with time. We proved indeed that regularly training the algorithm over time, and producing up-to-date sets of rules, are key components of the superior performance of the machine learning when it comes to stocks screening.

\section{Data}\label{sec: data}
The analyses in this paper are carried out on portfolios based on the investment universe defined by the capitalization-weighted MSCI World Index USD, that consists of the largest capitalization listed in the US, Canada, Western Europe, Japan, Australia, New Zealand, Hong Kong and Singapore. Portfolios are calculated in USD and net dividends are reinvested in the portfolio itself. Stock prices and dividends are taken from Thomson Reuters/Datastream. We reconstruct a proxy of the MSCI World Index by using end-of-month compositions as well as proxies for the US, Europe\footnote{Stocks in the MSCI World Index domiciled in Austria, Belgium, Denmark, Finland, France, Germany, Ireland, Italy, Netherlands, Norway, Portugal, Spain, Sweden, Switzerland and United Kingdom} and Asia Developed\footnote{Stocks in the MSCI World Index domiciled in Australia, Hong Kong, Japan, New Zealand and Singapore} benchmarks. We also consider sector portfolios derived from the MSCI World Index and the regional benchmarks by filtering on stocks that belong to the same sector: Consumer Staples (CS), Consumer Discretionary (CD), Energy (EN), Financials (FI), HealthCare (HC), Industrials (IN), Information Technology (IT), Materials (MA), Telecommunication Services (TL), Utilities (UT).\\
For each company in the investment universe, we collect ESG ratings from Sustainalytics\footnote{One of the largest provider of ESG ratings.}. An ESG rating is a comprehensive measure based on three pillars, Environment, Social and Governance, that assesses the strengths and weaknesses of a company along these three directions. The pillars are themselves based on a large set of specific indicators. For the purposes of this study, the composite ESG rating is the arithmetic average of the three ratings Environment (E), Social (S) and Governance (G), each of which is itself the combination of roughly 50 narrower indicators. Finally, for each company, we consider its relative Peer Group, which consists of all companies with a similar business, hence comparable from a sustainability point of view.

\section{The best-in-class approach}\label{sec: bic}

One of the most popular approaches to embed ESG criteria in the portfolio construction process is the so-called best-in-class approach. 
Given a threshold $x$, one excludes the stocks whose ESG ratings belong to the lowest $x$-quantile. The exclusion is usually carried across peer groups, i.e. groups of stocks with very similar characteristics. The reason behind this is twofold:\\
\begin{itemize}
\item Removing stocks with low ESG ratings within peer groups insures that the final economic mesh of the filtered universe remains similar to the initial investment universe.
\item ESG ratings have a structural, sector-driven bias which usually favor specific sectors (ex. IT or HealthCare sectors) while penalizing others (ex. Energy or Utilities). Given this bias, the filtering over peer groups makes comparisons of ESG ratings independent of the sectors. \\
\end{itemize}
For the purpose of this study, an ESG best-in-class portfolio derived from a capitalization-weighted portfolio removes, within each peer group, the stocks whose ratings belong to the lowest $x$-quantile. The portfolio is finally scaled to sum up to one.\\
This approach, quite popular among investors, should not be thought of as a way to enhance performance. As Tables \ref{tab: msci_world_bic}--\ref{tab: msci_asia_bic} show, ESG best-in-class filters applied to standard capitalization-weighted indexes do not bring outperformance.\\
Except for Europe and relatively low threshold levels, we find small but negative excess returns and negative information ratios for the ESG best-in-class portfolios over their benchmarks with almost unchanged risks. Although the approach does not create outperformance per se, it does not carry structural under-performance either. Optimistically, one could accept the fact that embedding ESG objectives in a portfolio does not significantly modify its risk/return profile.\\

\begin{table}[!h]
\scriptsize
  \floatsetup{floatrowsep=qquad, captionskip=4pt}
  \begin{floatrow}[2]
    \makegapedcells
    \ttabbox%
    {\hspace*{-1.35cm}\begin{tabularx}{0.58\textwidth}{c*{4}{>{\centering\arraybackslash}X}}
      \toprule
      & & \multicolumn{3}{c}{{ESG best-in-class}}\\
      \cmidrule(lr){2-5}
      & Bench.& 10\% & 30\% & 50\% \\
      \cmidrule(lr){2-2} \cmidrule(lr){3-5}
		Ann. Performance&10.07\%&10.01\%&9.93\%&9.51\%\\
		Ann. Volatility&13.34\%&13.31\%&13.44\%&13.82\%\\
		Sharpe Ratio&0.73&0.73&0.72&0.67\\
		Max. Drawdown&-21.91\%&-21.79\%&-22.02\%&-22.57\%\\
		Information Ratio&0&-0.27&-0.25&-0.41\\
      \bottomrule
      \end{tabularx}}
    {\caption{World Developed}
      \label{tab: msci_world_bic}}
    \hfill%
    \ttabbox%
    {\begin{tabularx}{0.58\textwidth}{c*{4}{>{\centering\arraybackslash}X}}
      \toprule
      & & \multicolumn{3}{c}{{ESG best-in-class}}\\
      \cmidrule(lr){2-5}
      & Bench.& 10\% & 30\% & 50\% \\
      \cmidrule(lr){2-2} \cmidrule(lr){3-5}
		Ann. Performance&13.45\%&13.25\%&13.46\%&13.2\%\\
		Ann. Volatility&14.61\%&14.54\%&14.49\%&14.4\%\\
		Sharpe Ratio&0.9&0.89&0.91&0.9\\
		Max. Drawdown&-18.99\%&-18.87\%&-18.71\%&-18.04\%\\
		Information Ratio&0&-0.71&0.02&-0.18\\      
      \bottomrule
      \end{tabularx}}
    {\caption{US}
      \label{tab: msci_usa_bic}}     
  \end{floatrow}
  \vspace*{1cm}
  \begin{floatrow}[2]
  \makegapedcells
    \ttabbox%
    {\hspace*{-1.35cm}\begin{tabularx}{0.58\textwidth}{c*{4}{>{\centering\arraybackslash}X}}
      \toprule
      & & \multicolumn{3}{c}{{ESG best-in-class}}\\
      \cmidrule(lr){2-5}
      & Bench.& 10\% & 30\% & 50\% \\
      \cmidrule(lr){2-2} \cmidrule(lr){3-5}
		Ann. Performance&6.37\%&6.55\%&6.47\%&6.31\%\\
		Ann. Volatility&19.25\%&19.19\%&19.19\%&19.29\%\\
		Sharpe Ratio&0.32&0.33&0.32&0.31\\
		Max. Drawdown&-30.25\%&-30.21\%&-30.2\%&-30.54\%\\
		Information Ratio&0&0.43&0.22&-0.11\\
      \bottomrule
      \end{tabularx}}
    {\caption{Europe}
      \label{tab: msci_europe_bic}}
    \hfill%
    \ttabbox%
    {\begin{tabularx}{0.58\textwidth}{c*{4}{>{\centering\arraybackslash}X}}
      \toprule
      & & \multicolumn{3}{c}{{ESG best-in-class}}\\
      \cmidrule(lr){2-5}
      & Bench.& 10\% & 30\% & 50\% \\
      \cmidrule(lr){2-2} \cmidrule(lr){3-5}
      Ann. Performance&6.83\%&6.71\%&6.41\%&5.75\%\\
	  Ann. Volatility&15.54\%&15.71\%&16\%&16.2\%\\
	  Sharpe Ratio&0.42&0.41&0.38&0.34\\
	  Max. Drawdown&-24.8\%&-24.95\%&-25.27\%&-25.8\%\\
	  Information Ratio&0&-0.21&-0.36&-0.46\\
      \bottomrule
      \end{tabularx}}
    {\caption{Asia}
      \label{tab: msci_asia_bic}}
  \end{floatrow}
  \parbox{\textwidth}{\hspace*{-1.35cm}$\, $\\}
  \parbox{1.14\textwidth}{\hspace*{-1.35cm}Key performance indicators of the MSCI World Index and three capitalization-weighted regional benchmarks, together with ESG best-}
  \parbox{1.14\textwidth}{\hspace*{-1.35cm}in-class filtered portfolios with different thresholds: 10\%, 30\% and 50\%. Data is shown in USD from August 2009 to March 2018. Source}
  \parbox{1.14\textwidth}{\hspace*{-1.35cm}MSCI, Datastream, Sustainalytics.}
 \end{table}%
$\, $\\
\noindent Our findings are not in contradiction with the large literature that finds positive links between ESG and financial performance. But the consistency and durability over time of the ESG factor has been questioned since the very beginning. \cite{Aupperle1985} find no significant relationship between social responsibility and corporate profitability, and similar results were obtained in \cite{CapelleBlancardMonjon2012} and \cite{HumphreyTan2014}. \cite{GriffinMahon1997} report that correlation between financial performance and social performance depends on the measure used to distinguish between high and low social performers.\\
Our results are more in line with \cite{RevelliViviani2015} for which \emph{"... the consideration of corporate social responsibility in stock market portfolios is neither a weakness nor a strength compared with conventional investments..."}. It should be noted that many fund managers and institutional investors surveys report that ESG is mostly looked as a risk mitigation tool in the first place \citep{VanDuuren2016}, and eventually as a performance driver at longer horizon. We share the optimistic view of Nobel prize-winning economist Robert Shiller for which both society and the financial community would find the use of socially responsible practices mutually beneficial \citep{Shiller2013}. At the same time, we also believe that short to mid term financial performance is at best lowly correlated to ESG ratings, at least for such broad investment universes as the MSCI World Index (which contains more than 1,600 companies). 
We can list several reasons for this:\\
\begin{enumerate}[(i)]
\item The investment universes are relatively large and the aggregated ESG ratings have too little a signal-to-noise ratio to allow for an efficient selection of outperforming stocks.
\item ESG ratings are global metrics that embrace environmental, social and governance criteria. As such, they may be too reductive and we may lose a significant amount of information from the single indicator to the aggregated scores. 
\item Granularity is key: As an example, it is likely that companies in specific sectors (ex. Energy) react differently to changes in the environment score (E) compared to the social score (S). 
\item In the search for a rational economic theory behind ESG, some argue that by divesting low ESG rated companies, investors raise their cost of capital and, in turns, the return these companies have to offer to attract new investors. As such, in the short run, they may show higher performances, but over time, the level of return they have to offer becomes unsustainable. Said otherwise, the action of divesting may take time to materialize in both investors' portfolios and low ESG rated companies (see for example \cite{AsnessBlog2017}).
\item The period considered in this study spans from the earlier stages of the recovery in 2009 to March 2018. Therefore, we are considering key performance indicators over a period of strong equity market, characterized by high returns and historically lower levels of volatility. This market regime can potentially affect the overall strength of ESG filtered portfolios.\\
\end{enumerate}
To illustrate item (iii), we consider sector portfolios derived from the MSCI World Index and from the three regional benchmarks (US, Europe, Asia) and we apply both ESG and single pillar E, S and G, 30\% best-in-class filtering. Tables \ref{tab: msci_world_sec}--\ref{tab: msci_asia_sec} collect the results. For the sake of simplicity, we only show annualized excess returns over the relative benchmark sector portfolios and information ratios.
\begin{table}[!h]
\tiny
  \floatsetup{floatrowsep=qquad, captionskip=4pt}
  \begin{floatrow}[2]
    \makegapedcells
    \ttabbox%
    {\hspace*{-0.95cm}\begin{tabularx}{0.5\textwidth}{c*{4}{>{\centering\arraybackslash}X}}
      \toprule
      & ESG& E & S & G \\
      \cmidrule(lr){2-5}
		CS&-0.26\% (-0.26)&-0.45\% (-0.47)&-0.65\% (-0.59)&\textbf{0.62\% (0.53)}\\
		CD&-0.33\% (-0.3)&-0.46\% (-0.46)&-0.65\% (-0.4)&-1.22\% (-0.65)\\
		EN&-0.24\% (-0.13)&-0.41\% (-0.19)&-0.26\% (-0.17)&\textbf{0.86\% (0.31)}\\
		FI&-0.51\% (-0.43)&-0.46\% (-0.42)&-0.87\% (-0.63)&\textbf{0.19\% (0.13)}\\
		HC&-0.39\% (-0.35)&-0.5\% (-0.3)&-0.47\% (-0.4)&-0.53\% (-0.43)\\
		IN&-0.31\% (-0.28)&-0.32\% (-0.27)&-0.32\% (-0.31)&-0.33\% (-0.19)\\
		IT&-0.13\% (-0.12)&-0.44\% (-0.45)&-1.53\% (-0.52)&\textbf{0.1\% (0.05)}\\
		MA&-0.09\% (-0.05)&-0.09\% (-0.03)&-0.17\% (-0.08)&\textbf{0.32\% (0.18)}\\
		TL&\textbf{1.26\% (0.64)}&\textbf{1.17\% (0.74)}&\textbf{0.15\% (0.06)}&\textbf{0.73\% (0.21)}\\
		UT&\textbf{0.16\% (0.1)}&-0.94\% (-0.39)&\textbf{0.72\% (0.43)}&\textbf{0.82\% (0.32)}\\
      \bottomrule
      \end{tabularx}}
    {\caption{World Developed}
      \label{tab: msci_world_sec}}
    \hfill%
    \ttabbox%
   {\hspace*{0.95cm}\begin{tabularx}{0.5\textwidth}{c*{4}{>{\centering\arraybackslash}X}}
      \toprule
      & ESG& E & S & G \\
      \cmidrule(lr){2-5}
		CS&-0.94\% (-0.84)&-0.46\% (-0.43)&-0.46\% (-0.32)&-0.26\% (-0.12)\\
		CD&-1.95\% (-0.82)&-0.48\% (-0.39)&-0.91\% (-0.55)&-2.92\% (-0.98)\\
		EN&-0.09\% (-0.06)&-0.66\% (-0.44)&\textbf{0.01\% (0.01)}&\textbf{0.55\% (0.16)}\\
		FI&-0.22\% (-0.19)&-0.11\% (-0.11)&-0.2\% (-0.17)&\textbf{1.02\% (0.67)}\\
		HC&-0.23\% (-0.22)&\textbf{0.19\% (0.15)}&-0.55\% (-0.5)&-0.58\% (-0.37)\\
		IN&\textbf{0.4\% (0.54)}&\textbf{0.25\% (0.32)}&\textbf{0.01\% (0.01)}&-0.6\% (-0.56)\\
		IT&-1.36\% (-0.96)&-0.7\% (-0.75)&-1.76\% (-0.51)&-0.37\% (-0.11)\\
		MA&\textbf{0.34\% (0.18)}&\textbf{0.72\% (0.32)}&-0.76\% (-0.32)&-0.14\% (-0.07)\\
		TL&-0.38\% (-0.37)&-0.32\% (-0.38)&-0.4\% (-0.2)&\textbf{1.32\% (0.28)}\\
		UT&\textbf{0.63\% (0.56)}&\textbf{0.47\% (0.38)}&\textbf{0.7\% (0.7)}&\textbf{0.13\% (0.12)}\\
      \bottomrule
      \end{tabularx}}
    {\caption{US}
      \label{tab: msci_usa_sec}}     
  \end{floatrow}
  \vspace*{1cm}
  \begin{floatrow}[2]
  \makegapedcells
    \ttabbox%
    {\hspace*{-0.95cm}\begin{tabularx}{0.5\textwidth}{c*{4}{>{\centering\arraybackslash}X}}
      \toprule
      & ESG& E & S & G \\
      \cmidrule(lr){2-5}
		CS&\textbf{0.16\% (0.14)}&\textbf{0.03\% (0.03)}&-0.01\% (-0.01)&\textbf{0.39\% (0.32)}\\
		CD&\textbf{0.51\% (0.32)}&\textbf{0.45\% (0.29)}&\textbf{0.18\% (0.14)}&\textbf{1.07\% (0.68)}\\
		EN&-1.2\% (-0.3)&-2.01\% (-0.38)&-0.5\% (-0.12)&-1.05\% (-0.27)\\
		FI&\textbf{0.31\% (0.18)}&\textbf{0.04\% (0.03)}&\textbf{0.33\% (0.17)}&-0.28\% (-0.21)\\
		HC&-0.38\% (-0.49)&-0.44\% (-0.55)&-0.53\% (-0.67)&-0.11\% (-0.18)\\
		IN&\textbf{0.03\% (0.03)}&\textbf{0.03\% (0.03)}&-0.39\% (-0.35)&-0.67\% (-0.44)\\
		IT&-0.63\% (-0.29)&-1.13\% (-0.49)&\textbf{0.01\% (0.08)}&-0.71\% (-0.43)\\
		MA&\textbf{0.12\% (0.03)}&\textbf{0.23\% (0.06)}&\textbf{0.51\% (0.14)}&\textbf{0.6\% (0.18)}\\
		TL&\textbf{1.81\% (0.67)}&\textbf{1.72\% (0.63)}&\textbf{1.82\% (0.62)}&\textbf{2.06\% (0.71)}\\
		UT&\textbf{1.79\% (0.54)}&\textbf{3.25\% (0.87)}&\textbf{0.09\% (0.03)}&\textbf{0.41\% (0.16)}\\
      \bottomrule
      \end{tabularx}}
    {\caption{Europe}
      \label{tab: msci_europe_sec}}
    \hfill%
   \ttabbox%
   {\hspace*{0.95cm}\begin{tabularx}{0.5\textwidth}{c*{4}{>{\centering\arraybackslash}X}}
      \toprule
      & ESG& E & S & G \\
      \cmidrule(lr){2-5}
        CS&\textbf{0.6\% (0.52)}&\textbf{0.37\% (0.12)}&-0.3\% (-0.28)&-0.18\% (-0.09)\\
		CD&-0.29\% (-0.19)&-0.11\% (-0.11)&\textbf{0.09\% (0.04)}&\textbf{0.75\% (0.37)}\\
		EN&\textbf{0.12\% (0.04)}&\textbf{0.34\% (0.11)}&\textbf{0.03\% (0.01)}&\textbf{0.25\% (0.1)}\\
		FI&-0.27\% (-0.16)&-0.09\% (-0.06)&-0.71\% (-0.46)&-0.08\% (-0.05)\\
		HC&\textbf{0.29\% (0.19)}&-0.67\% (-0.27)&\textbf{0.14\% (0.08)}&-0.03\% (-0.02)\\
		IN&-0.76\% (-0.39)&-0.41\% (-0.22)&-0.42\% (-0.22)&-0.61\% (-0.35)\\
		IT&-1.32\% (-0.59)&-0.92\% (-0.4)&-1.01\% (-0.43)&-0.96\% (-0.28)\\
		MA&-0.29\% (-0.26)&\textbf{0.04\% (0.03)}&-0.58\% (-0.44)&-0.07\% (-0.05)\\
		TL&\textbf{0.17\% (0.04)}&-0.18\% (-0.11)&-0.61\% (-0.09)&\textbf{1.83\% (0.24)}\\
		UT&\textbf{3.07\% (0.65)}&-3.57\% (-0.87)&\textbf{0.16\% (0.04)}&\textbf{2.6\% (0.55)}\\
      \bottomrule
      \end{tabularx}}
    {\caption{Asia}
      \label{tab: msci_asia_sec}}
  \end{floatrow}
  \parbox{\textwidth}{\hspace*{-0.95cm}$\, $\\}
  \parbox{1.04\textwidth}{\hspace*{-0.95cm}Annualized excess returns (Information ratios) between capitalization-weighted sector portfolios and their ESG best-in-class filtered versions for }
  \parbox{1.04\textwidth}{\hspace*{-0.95cm}the MSCI World Index and the derived regional benchmarks. In bold pairs sector/indicator for which the excess return is positive. Best-in-class}
  \parbox{1.04\textwidth}{\hspace*{-0.95cm}filters are performed with the ESG rating together with the single pillars Environment (E), Social (S) and Governance (G) ratings. Data is shown}
  \parbox{1.04\textwidth}{\hspace*{-0.95cm}in USD from August 2009 to March 2018. Source MSCI, Datastream, Sustainalytics.}
 \end{table}%

Overall, it is not straightforward to detect clear patterns between excess returns and ESG metrics conditionally to the regional benchmarks. But we can definitely detect specific triplets sector/region/metric that produce significant positive excess returns. Clearly, integrating ESG criteria in the Utilities sector enhances in-sample performances. But the right metric to use clearly depends on the geography: in the World Developed (Table \ref{tab: msci_world_sec}) the best excess return for the Utilities sector is achieved when one uses the Governance (G) score only at 0.82\%; in the US (Table \ref{tab: msci_usa_sec}) it is better to look at the composite ESG ratings which achieves 0.63\%. In Europe (Table \ref{tab: msci_europe_sec}) it is the Environment score (E) that obtains the best result with 3.25\% while in the Asia (Table \ref{tab: msci_asia_sec}) it is, once again, the composite ESG rating that achieves the highest excess return at 3.07\%.\\
$\, $\\
\noindent More generally, there is no sector nor metric for which the excess return of the best-in-class filtered sector achieves positive excess return in all the regions. Similarly, there is no region nor sector for which all metrics produce positive excess returns. Finally, no sector achieves positive excess returns across all regions and metrics. In other words, finding performance drivers when integrating ESG criteria in a best-in-class fashion is out of reach.

From Tables \ref{tab: msci_world_sec}--\ref{tab: msci_asia_sec}, only 12 out of 40 sector/metric portfolios in the World Developed region turn out to have positive excess return, and half of them are obtained when one considers the Governance (G) score. In the US we find positive excess returns in 14 out of 40, with no clear indication on the best metric to use. We notice though that all the metrics seem to work in the Utilities sector. 

In Europe, we count 25 out of 40 sector/metric pairs with positive excess returns. For 4 sectors (Consumer Discretionary, Materials, Telecommunication Services and Utilities) all metrics work accurately. In Asia, we have 16 out of 40 portfolios with positive excess returns with no clear patterns between sectors and metrics, except for the Energy sector for which all metrics produce positive excess returns, even if their magnitudes are relatively small.

\noindent In conclusion, our empirical findings confirm that simple ESG filtering does not bring extra performance. Overall, it rather behaves as a small drag. Given the short period we consider, and the market regime that equity markets have experienced since 2009, we share the view that ESG best-in-class integration is, most likely, neutral to financial performance. Nevertheless, our results highlight the fact that geographies and sectors do not react to ESG criteria in the same way.  But finding interesting and statistically significant patterns between ratings, pillars, their underlying narrow indicators (\emph{features}) and financial performances, for more than 150 indicators on more than 1,600 companies in the MSCI World Index, over roughly 10 year, is out of reach for both human and linear statistic tools. 

Next section introduces other techniques that can overcome this complexity and exploit this huge set of data.

\section{Machine learning}\label{sec: ml}

In this section we introduce a deterministic, easily understandable machine-learning prediction algorithm, aimed at finding consistent and statistically significant patterns between ESG ratings and financial performances. The algorithm explores a high-dimensional data-set of ESG granular indicators for all the companies in our investment universe. 

The goal of the algorithm, which falls in the category of supervised machine learning, is to predict the (conditional) excess return of each company over the benchmark, given the specific values taken by some of its ESG indicators (the features). Stated differently, the algorithm identifies regions in the high-dimensional space of ESG features that are statistically related to financial outperformance or underperformance. Features include raw and derived ESG indicators\footnote{For each raw indicator, as for example the environment score (E), we also look at the derived indicator relative to the peer-group and the sector. All of these transformations can potentially contain useful information. On the other side, the use of both raw and derived indicators rapidly increases the dimension of the feature space $d$.}, sector and country classifications, company's size and controversy indicators.

The regions are characterized by \emph{rules} in the form \emph{If-Then}, so that the algorithm finally consists of a set of such rules.
The \emph{If} statement is a list of conditions on the features $\x_t \in \Xset = \Xset_1\times\cdots\times \Xset_d$, where $\Xset_i$ is the set of possible outcomes of the feature $i$ and $d=447$ is the total number of features\footnote{We use 164 ESG raw indicators, from which we derive peer group and sector relative indicators and 3 valuation indicators. In total 164*3 + 3 = 495. From these indicators we remove 48 indicators for which either the sector or the peer group derived indicators are too close, or for which historical data is missing.}. Therefore, a rule defines a hyper-rectangle of $\Xset$. The \emph{Then} statement is the prediction of the 3-month forward excess return conditionally to the \emph{If} statement. Since the rules correspond to hyper-rectangles in the feature space, we finally obtain relatively simple and understandable regions. Furthermore, to avoid over-fitting, the algorithm only selects a finite number of such rules. At each time $t$, the predictions of each rule are aggregated into one prediction, $\hat{y}_t$, through convex combination.\\
\noindent The algorithm is calibrated (\emph{trained}) on the training set and the rules are used out-of-sample. The learning process works at two independent levels:
\begin{itemize}
\item At the end of year $N+1$ we train the algorithm on an expanded data-set of features and stock total returns which contains the data-set used at the end of year $N$ augmented of all the new observed data (features and stock total returns) from the end of year $N$ to the end of year $N+1$. To initialize the algorithm, we train it over 3 year of data (from 2009 to 2012). By expanding the data-set, the algorithm is able to access new data and explore new patterns, so that it can strengthen or nuance some rules that were previously discovered. \\
\item On a daily basis, the algorithm can update the weights used to aggregate each rule's prediction, by over-weighting rules with a good prediction rate and under-weighting the others. Therefore, following day predictions will benefit from the \emph{experience} the algorithm is gaining on the rules and their predictive power. The weight of each rule can be viewed as a confidence index.
Of course, this is possible because the algorithm is able to assess the goodness of its predictions by looking at the realized 3-month return.\\
\end{itemize}
\noindent To avoid threshold effects, we transform the final prediction for each stock into a score $S\in\{-1, 0, +1\}$, where $+1$ stands for significantly positive excess return prediction, $-1$ for negative prediction and $0$ for an uncertain prediction. The case where $S=0$ is usually related to stocks for which some of their ESG indicators would eventually signal financial outperformance, while other ESG indicators rather signal potential undeperformance. The picture is then nuanced, and the algorithm cannot make a precise prediction. This is a very common situation in finance, where different indicators can yield different forecasts, so that, in aggregate, the forecast turns out to be uninformative.

The learning process is divided into two steps. Following \cite{Nemirovski00} and \cite{Tsybakov03}, the training set $D_N$ at the end of year $N$ is divided into two sub-datasets: $D_n$ the learning set and $D_t$ the aggregation set, with $t >> n$ and $n+t=N$. The learning set $D_n$ is used to design and select the set of rules used by the algorithm to make predictions. The  aggregation set is used to fit the coefficients of the convex combination, in line with the \emph{expert aggregation theory} of \cite{PLG} and \cite{Stoltz10}.

\paragraph{Independent Suitable Rules}
Let $D_N = \left( (\x_1, y_1),\dots,(\x_N, y_N) \right) \in (\Xset \times \R)^N$ be the training set. Here $y_i$ denotes the 3-month return for some stock and $\x_i$ is the $d$-dimensional vector of its ESG features. The training set consists of a large but finite numbers of $(d+1)$-vectors spanning all stocks in the investment universe and all available dates. The training set $D_n\subseteq D_N$ includes the first $n$ data points in $D_N$ and $D_t = \left( (\x_{n+1}, y_{n+1}),\dots,(\x_N, y_N) \right)$, the order being induced by the time.

\begin{definition} 
	For any set $E \subset \Xset$, we define
	\begin{equation*}
		\mu(E, D_n):=\dfrac{\sum \limits_{i=1}^n y_i \ind{\x_i \in E}}{\sum \limits_{i=1}^n \ind{\x_i \in E}},
	\end{equation*}
here, by convention, $\frac00 = 0$.
\end{definition}
The set-valued map $\mu$ represents the conditional excess return of a stock over the benchmark, given that its ESG features $\x$ belong to $E$.

\begin{definition}\label{rule_def} 
Let $\r$ be a hyper-rectangle on $\Xset:\, \r =\prod_{k=1}^{d}I_k$ where each $I_k$ is an interval of $\Xset_k$.\\ A rule $f$ is a function defined on $\r \times \left(\Xset\times \R \right)^n$ as
	\begin{equation}
		f\left(\x, D_n\right) := \mu(\r, D_n)\, \forall \x\in\r
	\end{equation}
The hyper-rectangle $\r$ is called the \textbf{condition} and $\mu(\r, D_n)$ is called the \textbf{prediction} of the rule $f$. The event  $\{\x \in \r\}$ is called the \textbf{activation conditions} of the rule $f$.\\
\end{definition}
\noindent A rule $f$ is completely defined by its condition $\r$. So, with an abuse of notation, we do not distinguish between a rule and its condition. We define two crucial numbers for a rule:
\begin{definition} Let $f$ be a rule as in Definition \ref{rule_def} defined on $\r=\prod_{k=1}^{d}I_k$.
	\begin{enumerate}
		\item The \emph{number of activations} of $f$ in the sample $D_n$ is 
		\begin{equation*}
		n(\r, D_n) := \sum\nolimits_{j=1}^{n} \ind{\x_j \in \r}.
		\end{equation*}
		\item The \emph{complexity} of $f$ is 
		\begin{equation*}
		cp(\r)=d-\#\{1\le k\le d; I_k =\Xset_k\},
		\end{equation*}
	\end{enumerate}
\end{definition}

\noindent The algorithm does not consider all the possible rules, but only those with a given \emph{coverage} and \emph{significance}. We call these rules \emph{suitable}, and their definition is given below. 

\begin{definition}
	A rule $f$, defined on $\r$, is a \emph{suitable rule} for the training set $D_n$ if and only if it satisfies the two following conditions:
	\begin{enumerate}
		\item \textbf{Coverage condition.}
		\begin{equation}\label{cov_crit} 
			C_{min} \leq \dfrac{n(\r, D_n)}n \leq C_{max},
		\end{equation}
		with $C_{min}$ and $C_{max}$ are suitably chosen in the calibration step.
		\item \textbf{Significance condition.}
		\begin{equation}\label{sini_crit} 
			\left|  \mu(\r, D_n) -  \mu(\Xset, D_n)\right| \geq z(\r, D_n, \alpha),
		\end{equation}
		for a chosen $\alpha \in [0,1]$ and function $z$.
	\end{enumerate}
\end{definition}

\noindent The coverage condition \eqref{cov_crit} excludes rules that are activated only on small sets (i.e with a low coverage rate, $C_{min}$) and rules that are too obvious (i.e with a high coverage rate, $C_{max}$). The threshold in the significance condition~\eqref{sini_crit} is set such that the probability of falsely rejecting the null hypothesis $ \mu(\r, D_n)=\mu(\Xset, D_n)$ is less than $\alpha$. The parameter $\alpha$ permits to control the number of suitable rules. The higher $\alpha$, the higher the number of suitable rules. In what follows, we generate rules of complexity $c \geq 2$ by a \emph{suitable intersection} of rules
of complexity $1$ and rule of complexity $c-1$.
\begin{definition}\label{suitable_intersection}
	Two rules $f_i$ and $f_j$ defined on $\r_i$ and $\r_j$ respectively, form a \emph{suitable intersection} if and only if they satisfy the two following conditions:
	\begin{enumerate}
		\item \textbf{Intersection condition:}
		\begin{equation}\label{inter_crit} 
		\begin{split}
		& \r_i \cap \r_j \neq \emptyset,\\
		& n(\r_i \cap \r_j, D_n) \neq n(\r_i, D_n),\\
		& n(\r_i \cap \r_j, D_n) \neq n(\r_j, D_n)
		\end{split}
		\end{equation}
		\item \textbf{Complexity condition:}
		\begin{equation}\label{cp_crit}
		cp(\r_i \cap \r_j) = cp(\r_i) + cp(\r_j).
		\end{equation}
	\end{enumerate}
\end{definition}
\noindent The intersection condition \eqref{inter_crit} avoids adding a useless condition for a rule. In other words, to define a suitable intersection, $\r_i$ and $\r_j$ must not be satisfied by the same points in $D_n$. The complexity condition \eqref{cp_crit} means that $\r_i$ and $\r_j$ have no marginal index in common.

\paragraph{Designing Suitable Rules}

The design of suitable rules is made recursively on their complexity. It stops at a complexity $c$ if no rule is suitable or if the maximal complexity $c=cp_{max}$ is achieved.\\

\begin{description}
\item[Complexity $1$: ] The first step is to find suitable rules of complexity $1$. First notice that the complexity of evaluating all rules of complexity $1$ is $O(ndm^2)$. Rules of complexity $1$ are the base of the algorithm search heuristic. So all rules are considered and only suitable ones are kept, i.e rules that satisfied the coverage condition \eqref{cov_crit} and the significance condition \eqref{sini_crit}. Since rules are considered independently, the search can be parallelized. \\
\item[Complexity $c$: ] Among the  suitable rules of complexity $1$ and $c-1$, we select $M$ rules of each complexity ($1$ and $c-1$) according to a chosen criterion. Then it generates rules of complexity $c$ by pairwise \emph{suitable intersection} according to the Definition \ref{suitable_intersection}. The complexity of evaluating all rules of complexity $c$, obtained from their intersections, is $O(nM^2)$. Here again, since rules are considered independently, the evaluation can be parallelized. The parameter $M$ helps to control the computing time.
\end{description}

\paragraph{Selecting Suitable Rules} We select a subset $\S$ from all suitable rules which maximizes the gains expected from rule in $D_n$ and such as their conditions form a covering of $\Xset$.

\paragraph{Algorithm}
The calibration of the algorithm is structured in two parts: in the first one, it finds all suitable rules, and in the second one it retains only an optimal subset of it. To avoid threshold effects, overfitting and to manage the numerical complexity, we discretize each feature in $\Xset$ into $m$ classes with empirical quantiles (modalities)\footnote{Of course such procedure is performed only on real-valued features with more than $m$ different values. Categorical features are left unchanged.}. Thus, each modality of each variable covers about $100/m$ percent of the sample. In practice, $m$ must be inversely related to $d$: The higher the dimension of the problem, the smaller the number of modalities.
\newline

\noindent
The parameters of the algorithm are:
\begin{itemize}
	\item $m$, the sharpness of the discretization;
	\item $\alpha\in[0,1]$, which specifies the false rejecting rate of the test;
	\item $z$, the significance function of the test;
	\item $C_{max}$ and $C_{min}$ the coverage bounds;
	\item $cp_{max}$ the maximal complexity of a rule;
	\item and $M \in\mathbb N$, the number of rules of complexity $1$ and $c-1$ used to
	define the rules of complexity $c$.
\end{itemize}

\paragraph{Aggregation} Let $D_t = \left( (\x_{n+1}, y_{n+1}),\dots,(\x_{n+t}, y_{n+t}) \right) \in \left(\Xset, \R \right)^t$, where $n+t = N$ be the aggregation set and let $\S$ be the set of $R$ rules selected by the algorithm. At each time $t$, the predictions of each rule are aggregated into one prediction $\hat{y}_t$ as follows: 
\begin{equation}\label{y_hat}
	\hat{y}_{t} = \dfrac{\sum_{i=1}^{R} \pi_{i,t} f_i(\x_t, D_n)}{\sum_{i=1}^{R}\pi_{i,t}\ind{\x_t \in \r_i}}
\end{equation}
with $\pi_{i,1} = 1/R$. When the realized value $y_t$ is known, the weights $\pi_{.,t+1}$ are updated with the following formula
\begin{equation}\label{wexp}
	\pi_{i,t+1} = \pi_{i,t}\dfrac{e^{-\eta \ell(f_i(\x_t, D_n), y_t)}}{\sum_{i=1}^R e^{-\eta \ell(f_i(\x_t, D_n), y_t)}},
\end{equation}
with $\eta > 0$ and $\ell$ a convex loss function.

\begin{remark}
	One can notice that $f_i(\x_t, D_n)$ is not defined if $\x_t \notin \r_i$. In~\eqref{y_hat}, $\hat{y}_{t}$ is well defined for all $t$, since the set $\S$ is a covering of $\Xset$. In~\eqref{wexp} we follow the methodology of the \emph{sleeping expert aggregation} from \cite{Devaine13}.
\end{remark}
Once trained, the machine learning algorithm produces predictions of the excess returns which are transformed into a scores $S \in\{-1, 0, +1\}$, given the out-of-sample ESG features $\x_t$ for each company. Table \ref{tab: rules} shows some examples of rules taken from the learning process of the algorithm. The table lists three rules associated with positive predictions (opportunities) and five rules with negative predictions. Each rule consists of two features and two intervals. The "Relative To" properties indicate whether the feature must be calculated over all stocks in the universe (All), over a Sector, over a Peer Group, or whether we should look at the variations of the feature over time (Delta Score). 

Whenever the values taken by the features for a given company fall in the given intervals (we say that the stock activates the rule) the algorithm makes a prediction on its excess return. It is important to remark that we aggregate all the predictions, and we transform the final aggregated prediction into a score $\{-1, 0, +1\}$, so that in the end we mainly look at the sign of the prediction rather than at its magnitude. We also remark that, while the set of rules remains unchanged for one year (until the next learning process), the output of the rules can change over time, because raw indicators can change and also because the aggregated weights of the rules change over time.
\begin{table}[!h]
\scriptsize
\centering
\begin{tabular}{lccc}
\multicolumn{4}{c}{Opportunity Rules: Positive excess return}\\
\toprule
\multirow{2}{*}{Feature} & \multicolumn{1}{c}{Relative} & \multicolumn{1}{c}{Activation} & \multicolumn{1}{c}{Rule}\\ 
 & \multicolumn{1}{c}{To} & \multicolumn{1}{c}{Set} &  \multicolumn{1}{c}{Description}\\ 
\cline{1-4} \\[2pt]
Business Ethics Incidents     & \multicolumn{1}{c}{Sector} & \multicolumn{1}{c}{[5, 9]} & \multirow{2}{*}{\parbox{\rowHRul}{WHEN Business Ethics Incidents is high relative to sector AND
Board Remuneration Disclosure is high relative to sector THEN Opportunity}}\\[8pt]
Board Remuneration Disclosure & \multicolumn{1}{c}{Sector} & \multicolumn{1}{c}{[5, 9]}&    \\ [13pt]
\cline{1-4} \\[2pt]
Board Independence & \multicolumn{1}{c}{All} & \multicolumn{1}{c}{[9, 9]} &  \multirow{2}{*}{\parbox{\rowHRul}{WHEN Board Independence is at the maximum AND Board Remuneration Disclosure is high relative to sector THEN Opportunity}}\\[5pt]
Board Remuneration Disclosure & \multicolumn{1}{c}{Sector} & \multicolumn{1}{c}{[5, 9]} &   \\ [5pt]
\cline{1-4} \\[2pt]
Board Independence&\multicolumn{1}{c}{All}&\multicolumn{1}{c}{[9, 9]}&\multirow{2}{*}{\parbox{\rowHRul}{WHEN Board Independence is at the maximum AND Business Ethics Incidents is high relative to sector THEN Opportunity}}\\[5pt]
Business Ethics Incidents&\multicolumn{1}{c}{Sector}&\multicolumn{1}{c}{[5, 9]}&\\[5pt]
\cline{1-4} \\[5pt]
\multicolumn{4}{c}{Risk Rules: Negative excess return}\\
\toprule
\multirow{2}{*}{Feature} & \multicolumn{1}{c}{Relative} & \multicolumn{1}{c}{Activation} & \multicolumn{1}{c}{Rule}\\ 
 & \multicolumn{1}{c}{To} & \multicolumn{1}{c}{Set} &  \multicolumn{1}{c}{Description}\\ 
\cline{1-4} \\[2pt]
Verification of ESG Reporting&\multicolumn{1}{c}{Sector}&\multicolumn{1}{c}{[0, 7]}&\multirow{2}{*}{\parbox{\rowHRul}{WHEN Verification of ESG Reporting is not high relative to sector AND Board Remuneration Disclosure is low relative to sector THEN Risk}}\\[8pt]
Board Remuneration Disclosure&\multicolumn{1}{c}{Sector}&\multicolumn{1}{c}{[0, 4]}& \\ [13pt]
\cline{1-4} \\[2pt]
Quantitative Performance&\multicolumn{1}{c}{All}&\multicolumn{1}{c}{[5, 9]}&\multirow{2}{*}{\parbox{\rowHRul}{WHEN Quantitative Performance Score is high AND  Board Remuneration Disclosure is low relative to sector THEN Risk}}\\[5pt]
Board Remuneration Disclosure&\multicolumn{1}{c}{Sector}&\multicolumn{1}{c}{[0, 4]}& \\[5pt]
\cline{1-4} \\[2pt]
Verification of ESG Reporting&\multicolumn{1}{c}{All}&\multicolumn{1}{c}{[0, 6]}&\multirow{2}{*}{\parbox{\rowHRul}{WHEN Verification of ESG Reporting is not high AND Quantitative Performance Score is high THEN Risk}}\\[5pt]
Quantitative Performance&\multicolumn{1}{c}{All}&\multicolumn{1}{c}{[6, 9]}& \\[5pt]
\cline{1-4} \\[2pt]
Gender Diversity of Board&\multicolumn{1}{c}{Peer Group}&\multicolumn{1}{c}{[0, 8]}&\multirow{2}{*}{\parbox{\rowHRul}{WHEN Gender Diversity of Board is not high relative to peer group AND Employee Incidents is very low relative to peer group THEN Risk}}\\[5pt]
Employee Incidents&\multicolumn{1}{c}{Peer Group}&\multicolumn{1}{c}{[0, 2]}& \\[5pt]
\cline{1-4} \\[2pt]
Green Logistics Programs& \multicolumn{1}{c}{Delta Score} &\multicolumn{1}{c}{[0, 2]}&\multirow{2}{*}{\parbox{\rowHRul}{WHEN Green Logistics Programs Delta Score is very low AND Qualitative Performance Delta Score is very low THEN Risk}}\\[5pt]
Qualitative Performance & \multicolumn{1}{c}{Delta Score} &\multicolumn{1}{c}{[0, 2]}& \\[5pt]
\bottomrule
\end{tabular}
\caption{Some rules from the learning process of the algorithm at the end on 2012, 2013 and 2016. All features are discretized over 10 modalities (0 to 9) except for Qualitative Performance which is discretized over 6 modalities (0 to 5). High values for the features correspond to good ESG performance.}
\label{tab: rules}
\end{table}

\section{Machine learning application}\label{sec_ML_app}
We now test the predictive power of the machine learning algorithm developed in Section \ref{sec: ml} compared with the classical best-in-class approach. More precisely, we try to assess whether filtering stocks over scores derived from the algorithm outperforms the standard filtering over ESG ratings (best-in-class). 
For the sake of simplicity, we only present the World Developed universe and, among the strategies presented in Section \ref{sec: bic}, we only consider the 30\% best-in-class, as it is very close to what investors look at for their ESG portfolios. We recall that this strategy excludes, at each monthly review, the stocks whose ESG ratings are in the lower tercile within each peer group, and finally scale the weights so that their sum is one. To insure replicability of the portfolio, the ESG ratings are taken four days before the review date (which is end-of-month).\\
\noindent At the monthly review, we also build three portfolios based on the scores calculated with the machine learning algorithm, with the rules calculated at the end of the year that precedes the review:
\begin{description}
\item[Positive ML Screening:] The portfolio selects all stocks in the investment universe whose scores are $+1$. The weights are finally scaled up to sum to one (maintaining then the capitalization-weighting scheme of the benchmark)
\item[Positive ML Screening Sector Matched:] Same selection as for the Positive ML Screening portfolio, but the scaling of the weights is done in such a way that the final sector breakdown of the portfolio is matched to the benchmark's one.
\item[Negative ML Screening:] The portfolio selects all stocks in the investment universe whose scores are $-1$. The weights are finally scaled up to sum to one (maintaining then the capitalization-weighting scheme of the benchmark)
\end{description}
As before, the scores are taken four days before the review date. We consider the sector matched portfolio because the absolute screening usually introduces significant sector deviations with respect to the benchmark. Table \ref{tab: msci_world_ml} collects the main results for these portfolios since January 2013.\\
\begin{table}[!h]
\scriptsize
   \makegapedcells
    \begin{tabularx}{1\textwidth}{c*{5}{>{\centering\arraybackslash}X}}
     \toprule
      & & \multicolumn{3}{c}{{ML Screening}}& ESG best-in-class\\
      \cmidrule(lr){3-5} \cmidrule(lr){6-6}
      & \multirow{2}{*}{Bench.}& \multirow{2}{*}{Positive} & Positive &  \multirow{2}{*}{Negative} & \multirow{2}{*}{30\%}   \\      
      &                        &                           & Sect. Matched &  &  \\      
      \cmidrule(lr){2-2}\cmidrule(lr){3-5} \cmidrule(lr){6-6}
		Ann. Performance&10.32\%&13.07\%&11.66\%&8.31\%&10.13\%\\
Ann. Volatility&10.50\%&11.14\%&10.96\%&10.95\%&10.57\%\\
Sharpe Ratio&0.94&1.14&1.03&0.72&0.92\\
Max. Drawdown&-18.07\%&-14.99\%&-16.46\%&-22.47\%&-17.91\%\\
Information Ratio&-&1.01&0.58&-0.54&-0.32\\
Ann. Alpha&-&2.47\%&1.15\%&-1.81\%&-0.24\%\\
      \bottomrule
      \end{tabularx}
    \caption{Key performance indicators of the MSCI World Index (Bench.), the capitalization-weighted selection filtered over positive scores from the ML algorithm, the one with the sector allocation matched to the benchmark, the one screened over negative scores and the 30\% ESG best-in-class filtered portfolios. Data is shown in USD from January 2013 to March 2018. Source MSCI, Datastream, Sustainalytics.}
    \label{tab: msci_world_ml}
\end{table}
$\, $\\
\noindent Although we recognize that the period over which we can test the machine learning algorithm is relatively short (five years and three months), the results we obtain contain some interesting insights. 
First of all, the Positive ML Screening outperforms all the other portfolios: by 2.76\% the benchmark on an annualized basis, by 2.94\% the ESG best-in-class portfolio and by 4.77\% the Negative ML Screening. And while the realized annual volatilities remain in the range 10.50\% to 11.14\%, there are significant differences in the realized maximum drawdowns: the Negative ML Screening shows a -22.47\% loss from its peak, while the Positive ML Screening loss from its peak accounts for -14.99\%.\\
\begin{figure}[h!]
\centering
\includegraphics[width=0.8\textwidth]{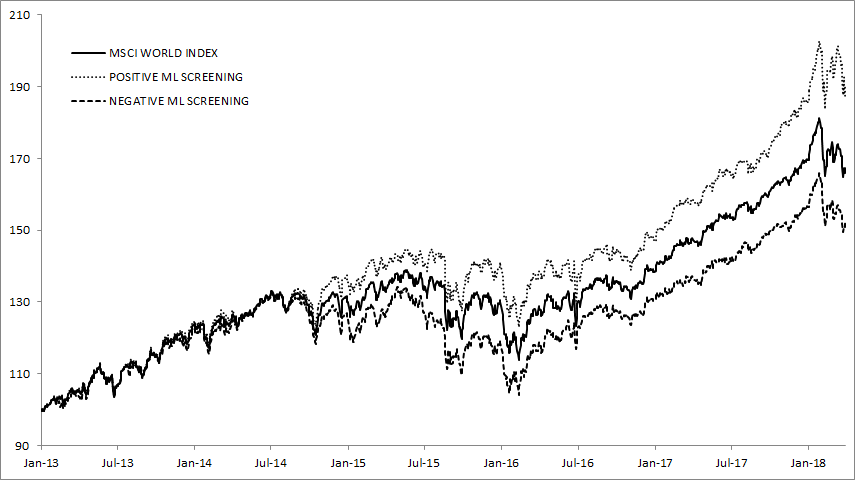}
\captionsetup{margin={40pt,48pt}}
\caption{Simulated strategy levels for the benchmark MSCI World Index, the capitalization-weighted selection filtered over positive scores from the ML algorithm (Positive ML Screening) and the one screened over negative scores (Negative ML Screening). Data in USD from January 2013 to March 2018. Base level = 100. Source MSCI, Datastream, Sustainalytics.}
\label{fig: Generic_measures}
\end{figure}
$\, $\\
\noindent These two combined results show that the machine learning algorithm is clearly able to distinguish between opportunity stocks (the ones with positive scores) from risky stocks (negative scores). Figure \ref{fig: Generic_measures} shows the historical behavior of these two portfolios and the benchmark. We notice that the Positive ML Screening outperforms the Negative ML Screening over time, with the benchmark in between. Furthermore, in years when the benchmark shows very high performances with very low volatility, typically in bull market regimes, the differences between the two strategies are less pronounced. On the contrary, when the market is in bear regimes or it does not have clear trend, the Positive ML Screening clearly outperforms its negative counterpart, as shown in Table \ref{tab: msci_world_yyperf}.
\begin{table}[!h]
\scriptsize
   \makegapedcells
    \begin{tabularx}{1\textwidth}{c*{5}{>{\centering\arraybackslash}X}}
     \toprule
      & & &\multicolumn{3}{c}{Excess Return}\\
      & & \multicolumn{3}{c}{ML Screening}& ESG best-in-class\\
      \cmidrule(lr){3-5} \cmidrule(lr){6-6}
      \multirow{2}{*}{Year}& \multirow{2}{*}{Bench.}& \multirow{2}{*}{Positive} & Positive &  \multirow{2}{*}{Negative} & \multirow{2}{*}{30\%}   \\      
      &                        &                           & Sect. Matched &  &  \\      
      \cmidrule(lr){2-2}\cmidrule(lr){3-5} \cmidrule(lr){6-6}
		2013&23.95\%&0.72\%&0.12\%&-1.39\%&-0.14\%\\
		2014&4.97\%&3.88\%&3.65\%&-2.75\%&-0.17\%\\
		2015&-0.89\%&3.79\%&1.89\%&-5.3\%&0.07\%\\
		2016&7.4\%&-2.14\%&-1.91\%&3.73\%&-0.44\%\\
		2017&22.44\%&3.82\%&0.61\%&-2.57\%&-0.02\%\\
		2018&-1.37\%&3.92\%&2.24\%&-1.63\%&-0.24\%\\
      \bottomrule
      \end{tabularx}
    \caption{Calendar year performances for the MSCI World Index (Bench.) and the excess returns for the Positive, Positive Sect. Matched and Negative ML Screening as well as for the ESG 30\% best-in-class portfolio. Data is shown in USD from January 2013 to March 2018. Source MSCI, Datastream, Sustainalytics.}
    \label{tab: msci_world_yyperf}
\end{table}
$\, $\\
\noindent In years when the benchmark performance is very significant (2013 or 2017), the Positive ML Screening is still able to achieve some outperformance, but the spread with the Negative ML Screening is  somehow lower than years when the market performance is negative or low (2014, 2015 and, most recently, 2018). 

Interestingly, the excess return of the sector matched version is also positive, even if lower in magnitude when compared to the Positive ML Screening. By neutralizing the sector component (because matched), the outperformance essentially comes from the stock picking.  

For the Negative ML Screening, the excess return is always negative except for 2016. Finally,  the best-in-class portfolio shows almost systematically small but negative excess returns, except in 2015 when it managed to outperform by 0.07\%. Once again, our findings confirm the fact that for very large and diversified universes, the simple ESG filtering does not bring alpha, although it does not significantly reduce the performance with the best-in-class approach.\\

\begin{figure}[!h]
\centering
        \begin{subfigure}[b]{0.475\textwidth}
            \centering
            \includegraphics[width=\textwidth]{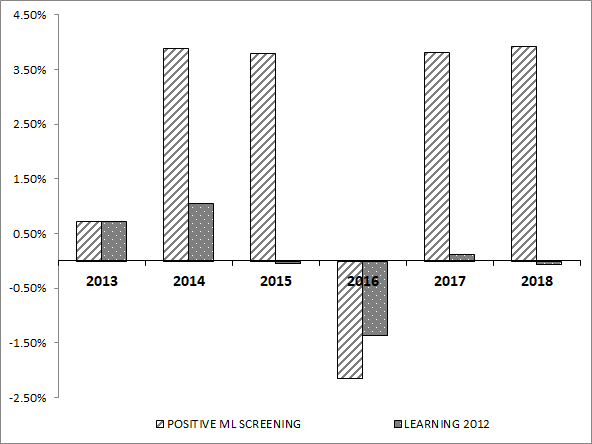}
            \caption{\scriptsize{Learning end of 2012}}   
        \end{subfigure}
        \hfill
        \begin{subfigure}[b]{0.475\textwidth}  
            \centering 
            \includegraphics[width=\textwidth]{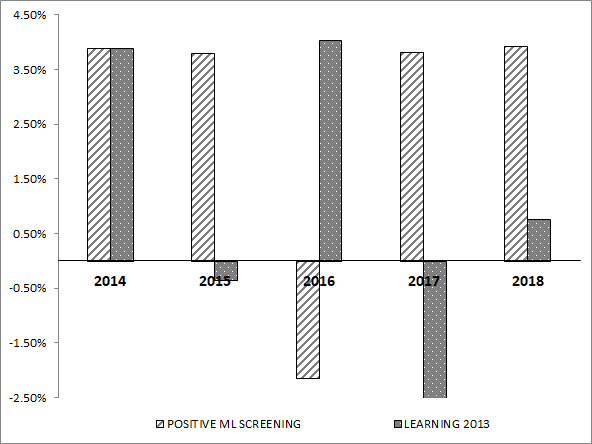}
            \caption{\scriptsize{Learning end of 2013}}  
        \end{subfigure}
        \vskip\baselineskip
        \begin{subfigure}[b]{0.475\textwidth}   
            \centering 
            \includegraphics[width=\textwidth]{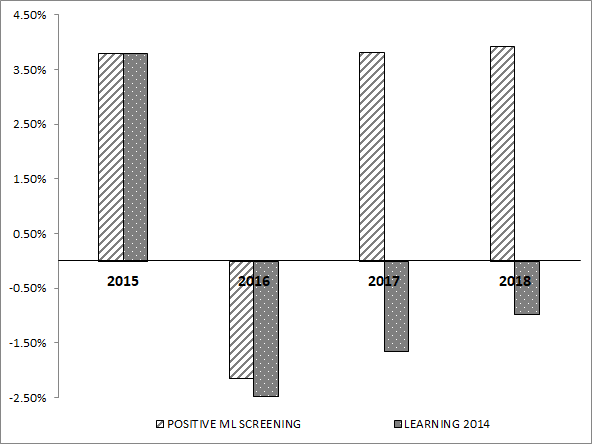}
            \caption{\scriptsize{Learning end of 2014}}    
        \end{subfigure}
        \quad
        \begin{subfigure}[b]{0.475\textwidth}   
            \centering 
            \includegraphics[width=\textwidth]{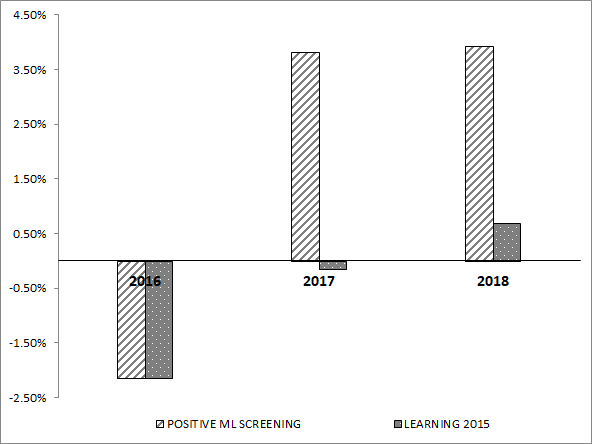}
            \caption{\scriptsize{Learning end of 2015}}   
        \end{subfigure}
\caption{\scriptsize{Calendar excess returns of the Positive ML Screening and the four portfolios LEARNING 2012, LEARNING 2013, LEARNING 2014 and LEARNING 2015 over the MSCI World Index. Data is shown in USD from January 2013 to March 2018. Source MSCI, Datastream, Sustainalytics.}} 
        \label{fig:learnings}
\end{figure}

\paragraph{The effects of learning} The machine learning algorithm is initially trained over three years of data and then yearly updated. During these regular updates, the algorithm learns from the new flow of data it can access: It can test its rules to confirm, nuance or remove some of them, and selects new rules linked to statistically significant patterns. This learning process is key in the final performance of the model (and for the Positive ML Screening portfolio built upon it). To measure this effect, we form four portfolios named LEARNING Y, where Y = 2012, 2013, 2014, 2015 as follows:

\begin{itemize}
\item For each year Y, we consider the set of rules related to the learning at the end of the year Y.
\item We calculate the scores for all stocks in the universe from the end of year Y to March 2018 with this set of rules.
\item LEARNING Y is built as Positive ML Screening, except that the underlying scores are calculated with the same, not updated set of rules calibrated at the end of year Y.
\end{itemize}
Said differently, LEARNING Y uses a unique, static set of rules that is never updated (no learning). By construction, the portfolios Positive ML Screening and LEARNING Y coincide over the period $\textit{January, 1}^{st},\, Y+1$ to $\textit{December, 31}^{st},\, Y+1$, because, over this period, they use the same set of rules (hence the same scores) to screen the investment universe. Figure \ref{fig:learnings} shows the calendar excess returns of these portfolios together with the Positive ML Screening portfolio over the benchmark MSCI World Index. \\
$\, $\\
\noindent Since we only show out-of-sample results, the time frame of each LEARNING Y portfolio is different. In the majority of cases, we see that Positive ML Screening outperforms the LEARNING Y portfolios after the first year (since they are the same on the first year). Indeed, the excess return for the LEARNING Y portfolios usually shrinks to zero and becomes even negative over time. 

In other words, the predictive power of the scores vanishes over time, so that it is important to train the algorithm on the new observed data to update the set of rules.
\begin{figure}[!h]
\centering
\begin{subfigure}[b]{0.475\textwidth}
  \centering
  \includegraphics[width=\textwidth]{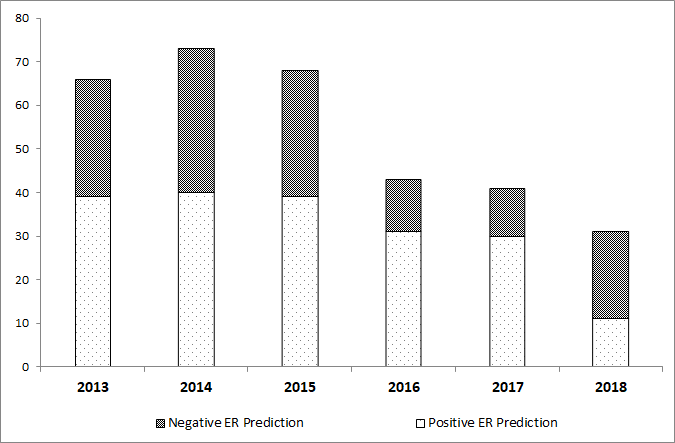}
  \caption{\scriptsize{Positive vs. Negative}}   
\end{subfigure}
\hfill
\begin{subfigure}[b]{0.475\textwidth}  
 \centering 
 \includegraphics[width=\textwidth]{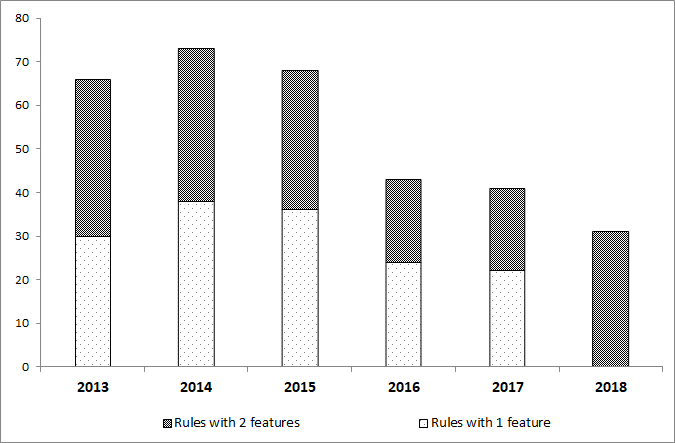}
 \caption{\scriptsize{Simple vs. Complex}}  
\end{subfigure}
\caption{\scriptsize{Number of rules at each update of the algorithm: (a) the split between rules that predict positive or negative excess returns (ER); (b) the split between rules that make use of one feature (Simple) or two features (Complex).}} 
\label{fig:rules}
\end{figure}
$\, $\\
\noindent The number of rules used by the algorithm changes over time: as shown in Panel (a) of Figure \ref{fig:rules}, this number evolves in the range $[31,73]$ with the split between positive rules (i.e. rules related to positive predictions of the excess return) and negative ones also changing over time. Interestingly, the number of rules related to negative excess return increased from 12 in 2016 to 20 in the latest 2018 learning. Panel (b) of Figure \ref{fig:rules} shows the same number of rules split between simple rules (i.e. those that only make use of one feature) and complex rules (i.e. those that use two features, as the examples shown in Table \ref{tab: rules}). Both Figures \ref{fig:learnings}-\ref{fig:rules} suggest that to extract alpha from the ESG features, one needs to regularly update the algorithm, and consider newly created set of rules to detect patterns between ESG profiles and financial performances.

\section{Conclusion}

The last few years have seen an increasing interest toward ESG investing and the integration of socially responsible principles at the portfolio construction level. Managers and investors are asked to complement pure financial objectives with extra-financial ones.

Our study brings some new ideas and insights onto the way investors could achieve ESG objectives in their investments. The literature on the theme is mixed: Initial studies were mostly skeptical on the benefit of ESG integration into the portfolio. Over time the mindset has evolved, and several studies have empirically proved that ESG integration in the portfolio does not lower performances. Most recently, the financial literature has gone one step further and claim that, indeed, ESG integration is a way to extract alpha or, at least, to reduce risks. 

We do recognize the need for serious integration of ESG objectives alongside with classic financial ones, and that there exists an economic link between the ESG profile of a company and its financial performances over the long run. Nevertheless, we tend to agree with the pioneers of ESG research for which, at best, ESG integration does not significantly degrade financial performances, especially for large and diversified investment universes. 

Because ESG profiles can impact financial performances in a non-linear way, and the impact can depend on the sector, the country or other specific characteristics of each company, we designed and implemented a sophisticated machine learning algorithm that identifies patterns between ESG profiles and performances, statistically robust across the universe and over time. 

The algorithm produces a set of rules, each rule identifying a region in the high-dimensional space of the ESG features, conditionally on which we can make a prediction on the stock's excess return. All the predictions are finally aggregated and transformed into a score taking values in $\{-1, 0, +1\}$, so that in the end we effectively look at the sign of the excess return rather than its magnitude.

With this algorithm, trained over time to keep it updated, we empirically proved that the link between ESG profiles and financial performances exists, but can only be accessed with non-linear techniques. Indeed a simple strategy that selects stocks whose scores are positive significantly outperforms the well known ESG best-in-class approach.\\
\vspace{1cm}

\bibliographystyle{chicago}

\end{document}